\documentclass[a4paper,12pt]{article}

\def\a{\alpha}

\def\bb{\begin{equation}}
\def\ee{\end{equation}}

\begin{document}

\begin{center}
{\huge Multidimensional integrable boundary problems}

\vskip 0.2cm

{Habibullin I.T. (Ufa, Institute of mathematics of RAS, Russia})

{e-mail: ihabib@imat.rb.ru}
\end{center}
\vskip 0.2cm

The article deals with the problem of finding integrable boundary
conditions for multi-dimensional equations. The research is
stimulated by the following problem, posed by Richard Ward: Find
discrete versions of the well known generalized Toda chains,
corresponding to the Lie algebras of finite growth (see R.Ward,
\cite{war}). Really, one has to find cutting off constraint for
the infinite discrete Toda chain
\begin{equation}
e^{f_{uv}-f_u-f_v+f}=\frac{1+he^{f_u^k-f_v}}{1+he^{f_u-f_v^{-k}}},
\label{dtoda}
\end{equation}
where $f=f(u,v,k)$ -- is the field variable, and sub- and
super-indices denotes shifts in the corresponding arguments:
$f_u=f(u+1,v,k)$, $f_v=f(u,v+1,k)$, $f^k=f(u,v,k+1)$,
$f^{-k}=f(u,v,k-1)$.

But in the paper we study the other well known model -- the Hirota
equation
\begin{equation}
T_uT_v-TT_{uv}=T^k_vT^{-k}_u\label{hiro},
\end{equation}
which is also a discrete version of the Toda chain. However the
algorithm proposed can be applied to any kind (discrete and
continuous \cite{hg}) of integrable equations with 3 independent
variables, including the Ward's problem. Cutting off constraint
for Hirota's equation are found corresponding to the generalized
Toda chains of the series $B_n$ and $C_n$.

We tried also to answer the question: What is the Lax pair of the
boundary value problem? It differs from that of the equation. Our
observation is that boundary condition put on the field variables
generates some boundary conditions for the eigenfunctions
(solutions to the linear Lax system). As a rule one has to take
two different Lax pairs with the eigenfunctions related to each
other at the border points. The best illustration to this idea
provides the example of the Kadomtsev-Petviashvili (KP) equation
below in the section 4. It can be checked by direct computation
that the boundary problem (\ref{kp1}), (\ref{bc010}) admits the
Lax pair (\ref{xv}), (\ref{vbc}), (\ref{tv}). In other words the
boundary problem (\ref{kp1}), (\ref{bc010}) is a consistency
condition of the linear boundary value problem (\ref{xv}),
(\ref{vbc}) with the time evolution (\ref{tv}). It is remarkable
that passing in the aforementioned example from the
eigenfunctions to the Gel'fand-Levitan-Marchenko kernel leads to
a linear boundary value problem which is solved by separation of
variables.

\section{Algorithm of searching for BC's}

How to extract the integrable BC's by using only the Lax
representation of the equation?

To answer the question let us concentrate on the example of the
Toda chain \bb \label{toda} u_{xt} (n) = \exp\{ u\left( {n - 1}
\right) - u\left( {n} \right)\} - \exp\{ u\left( {n} \right) -
u\left( {n + 1} \right)\} , \ee Its Lax pair is defined by
Laplace transformations for the following hyperbolic equation
\begin{eqnarray}
&&\label{todaxt} \phi _{xt} \left( {n} \right) =- u_{x} \left( {n}
\right)\phi _{t} \left( {n} \right) -e^{ u\left( {n - 1} \right) -
u\left( {n} \right)} \phi \left( {n}
\right),\\
&&\label{todan} \phi \left( {n + 1} \right) = \left( {D_{x} +
u_{x}
\left( {n} \right)} \right)\phi \left( {n} \right),\\
&&\label{todat}\phi_t(n)=-e^{u(n-1)-u(n)}\phi(n-1).
\end{eqnarray}
Take one more Lax pair of the same chain essentially different
from the above one
\begin{eqnarray}
&&\label{todaxt2} \psi _{xt} \left( {n} \right)=- u_{t} \left( {n}
\right)\psi _{x} \left( {n} \right) - e^{u\left( {n - 1}
\right) - u\left( {n} \right)} \psi \left( {n} \right),\\
&&\label{todan2} \psi \left( {n + 1} \right) = \left( {D_{t} +
u_{t} \left( {n} \right)} \right)\psi \left( {n} \right),\\
&&\label{todax2}\psi_x(n)=-e^{u(n-1)-u(n)}\psi(n-1).
\end{eqnarray}

Proposition 1 (see \cite{hg}). Let the operator of the form
$M=aD^3_x+bD^2_x+cD_x+d$ exist such that for $n=0$ for each
solution $\psi$ of the equation (\ref{todaxt2}) the function
$\phi=M\psi$ solves the equation (\ref{todaxt}). Then the field
variables $u(0), u(-1),u(-2)$ are connected by one of the
constraint
\begin{eqnarray}
&1)& e^{u(-1)}=0, \nonumber \\ &2)& u(-1)=0, \nonumber\\
&3)&u(-1)=-u(0),
\label{bct}\\&4)&u_x(-1)=-u_t(0)e^{-u(0)-u(-1)},\nonumber\\
&5)&e^{u(-2)-u(-1)}(u_x(-2)+u_x(-1))=\nonumber\\
&&=e^{u(-1)-u(0)} (6u_x(-1)-2u_x(0)). \nonumber
\end{eqnarray}
The corresponding differential operators $M$ are of the form
\begin{eqnarray}
&1)&M_1=a_0e^{u}(D^3_x+2u_xD^2_x+(u_{xx}+u_x^2)D_x)+\nonumber\\
&&+ b_0e^u(D^2_x+u_xD_x)+ c_0e^u D_x,\nonumber
\\ &2)& M_2=e^ {u}D^2_x+u_xe^ {u}D_x, \nonumber\\
&3)&M_3=e^ {u}D_x, \label{ot}\\
&4)&M_4=e^ {u}D^2_x+u_xe^ {u}D_x+e^ {-u},\nonumber\\
&5)&M_5=e^ {u}D^3_x+2u_xe^ {u}D_x^2+\nonumber\\
&&+(u_{xx}-u_{xx}(-1)+u_x^2- u_{x}^2(-1))e^ {u}D_x.\nonumber
\end{eqnarray}
where $a_0$, $b_0$, $c_0$ -- are real constant parameters, and
$u=u(0)$.

Conclusion: Under integrable boundary conditions two essentially
different Lax pairs (which are not related to each other by gauge
transform) become conjugate along the border. The conjugation
relation is nothing else but the corresponding boundary condition
for the eigenfunions.

\section{Boundary conditions for the KP equation
consistent with the Lax pair}

Apply the method suggested above to the KP equation
\begin{eqnarray}\label{kp}
u_{t}  + u_{xxx} - 6uu_{x}& =&-3\a^2w_{y},\\
w_x&=&u_{y}.\nonumber
\end{eqnarray}
Remind the Lax pair for it which is defined as the following
system of the linear equations
\begin{eqnarray}\label{xy}
&&\phi _{xx} = \alpha\phi _{y} + u\phi,\\
&&\label{xt} \phi _{t}=-4\phi _{xxx} + 6u\phi _{x} +3(u_x+\alpha
w)\phi.
\end{eqnarray}
The dual pair is of the form:
\begin{eqnarray}\label{xy2}
&&\tilde\phi _{xx} =-\alpha\tilde\phi _{y} + u\tilde\phi,\\
&&\label{xt2} \tilde\phi _{t}=-4\tilde\phi _{xxx} + 6u\tilde\phi
_{x} +3(u_x-\alpha w)\tilde\phi.
\end{eqnarray}

{\bf Lemma 1}. Solution $u=u(x,y,t)$ to the KP equation satisfies
the boundary condition $w|_{y=0}=0$ (or the boundary condition
$u_x+\alpha w|_{y=0}=0$) iff for each solution $\phi=\phi(x,y,t)$
of the equation (\ref{xt}) the function $\tilde\phi=\phi|_{y=0}$
(respectively the function $\tilde\phi=D_x\phi|_{y=0}$) solves
the equation (\ref{xt2}) for $y=0$.

\section{Zakharov-Shabat dressing and the boundary problem}

Let $u_0\equiv0$ be a starting solution to the KP and let
$\phi_0=\phi_0(x,y,t)$ be a starting eigenfunction. Find a new
(dressed) eigenfunction $\phi=\phi(x,y,t)$, corresponding to the
new (dressed) potential $u$. To this end we use the following
integral operator with Volterra kernel $K=K(x,z,y,t)$ (see
\cite{zs}):
\begin{eqnarray}\label{phi}&\phi(x,y,t)&=\phi_0(x,y,t)+\nonumber\\
&&+\int_{-\infty}^x K(x,z',y,t)\phi_0(z',y,t)dz'\end{eqnarray}
Its kernel $K$ satisfies the Gel'fand-Levitan-Marchenko equation
\begin{eqnarray}\label{mar}&&K(x,z,y,t)+F(x,z,y,t)+\nonumber\\
&&+\int_{-\infty}^x K(x,z',y,t)F(z',z,y,t)dz'=0.\end{eqnarray}
The kernel of the Gel'fand-Levitan-Marchenko equation $F$
satisfies in turn a system of linear differential equations with
constant coefficients
\begin{eqnarray}\label{Fy}
&&\alpha\frac{\partial F}{\partial y}- \frac{\partial^2
F}{\partial x^2}
+\frac{\partial^2 F}{\partial z^2}=0,\\
&&\label{Ft}\frac{\partial F}{\partial t}+4(\frac{\partial^3
F}{\partial x^3} +\frac{\partial^3 F}{\partial z^3})=0.
\end{eqnarray}
The searched solution $u$ of the nonlinear equation is defined
through the kernel $K$ by means of the formula
\bb\label{u}u(x,y,t)=2\frac{\partial}{\partial x}K(x,x,y,t).\ee
Formulate now the dressing method based on the other Lax pair,
then the system (\ref{Fy})-(\ref{Ft}) is replaced by the
following system
\begin{eqnarray}\label{Fy2}
&&-\alpha\frac{\partial\tilde{F}}{\partial y}- \frac{\partial^2
\tilde{F}}{\partial x^2}
+\frac{\partial^2 \tilde{F}}{\partial z^2}=0,\\
&&\label{Ft2}\frac{\partial \tilde{F}}{\partial
t}+4(\frac{\partial^3 \tilde{F}}{\partial x^3} +\frac{\partial^3
\tilde{F}}{\partial z^3})=0,
\end{eqnarray}
and the kernel $\tilde K$ of the transformation operator, which
converts the starting eigenfunction $\tilde{\phi}_0$ into an
eigenfunction $\tilde{\phi}$ of the new associated equation
\bb\label{tphi}\tilde{\phi}(x,y,t)=\tilde{\phi}_0(x,y,t)
+\int_{-\infty}^x \tilde{K}(x,z',y,t)\tilde{\phi}_0(z',y,t)dz'\ee
will solve the similar Gel'fand-Levitan-Marchenko equation
\begin{eqnarray}\label{mar2}&&\tilde{K}(x,z,y,t)+\tilde{F}(x,z,y,t)
+\nonumber\\
&&+\int_{-\infty}^x\tilde{K}(x,z',y,t)\tilde{F}(z',z,y,t)dz'=0.
\end{eqnarray}

{\bf Theorem} (see \cite{hg}). Involution \bb\label{iF}
\frac{\partial F}{\partial x}(x,z,0,t) =-\frac{\partial \tilde
F}{\partial z}(x,z,0,t)\ee corresponds to the boundary condition
\bb u_x+\alpha w|_{y=0}=0,\label{bc}\ee and the involution
\bb\label{iF0} F(x,z,0,t) = \tilde F(x,z,0,t),\ee corresponds,
respectively to the boundary condition \bb
w|_{y=0}=0.\label{bc0}\ee

Consider in more details the KP2 equation, putting in all
formulae above $\alpha=1$. Let the Gel'fand-Levitan-Marchenko
equation have the degenerate kernel \bb\label{4F}
F(x,z,y,t)=\sum^N_{n=1} c_n \exp((q_n^2-p_n^2)y
-4(q_n^3+p_n^3)t+q_nz+p_nx). \ee Then evidently \bb\label{4tF}
\tilde F(x,z,y,t)=\sum^N_{n=1} c_n \exp((q_n^2-p_n^2)y
-4(q_n^3+p_n^3)t+q_nx+p_nz). \ee Suppose in addition that the
kernels $F$ and $\tilde F$ are related by the involution
(\ref{iF}), which can be represented as $$\sum^N_{n=1} c_np_n
\exp\{q_nz+p_nx-4(q_n^3+p_n^3)t\}=$$ $$=-\sum^N_{n=1} c_n
p_n\exp\{q_nx+p_nz-4(q_n^3+p_n^3)t\}.$$ It implies immediately
that \bb\label{ic}\displaystyle {p_n=q_{N-n}},\quad
c_np_n=-c_{N-n}q_n.\ee Thus, under constraint (\ref{ic}) the
solution of the KP equation given by the formulae \bb\label{det}
u(x,y,t)=-2\frac{\partial^2}{\partial x^2}\ln\det A,\ee where $A$
is a matrix with the entries \bb\label{A}
A_{nm}=\delta_{nm}+\frac{c_n}{p_n+q_m}e^{((q_n^2-p_n^2)y
-4(q_n^3+p_n^3)t+(p_n+q_m)x)}, \ee satisfies the boundary
condition (\ref{bc}), which is $u_x+w|_{y=0}=0$ for the KP2. Its
differential consequence with respect to $x$ looks more
illustrative: $u_{xx}+u_y|_{y=0}=0.$

\section{Boundary problem on the stripe $0<y<1$}

Consider the boundary problem on the stripe $0<y<1$,
\begin{eqnarray}\label{kp1}
u_{t}  + u_{xxx} - 6uu_{x}& =&-3\a^2w_{y},\\
w_x&=&u_{y},\nonumber
\end{eqnarray}
\begin{equation}\label{bc010}
w|_{y=0}=0, \qquad w|_{y=1}=0,
\end{equation}
which is equivalent to a compounded Lax pair. Eigenfunction
solves the boundary problem
\begin{equation}\label{xv}
D^2_x\pmatrix{\phi\cr\psi}=\pmatrix{\alpha&0\cr0&-\alpha}D_y
\pmatrix{\phi\cr\psi}+u \pmatrix{\phi\cr\psi}
\end{equation}
with boundary conditions imposed along the same direct lines
$y=0$ and $y=1$
\begin{equation}\label{vbc}
\phi-\psi|_{y=0}=0,\qquad \phi-\psi|_{y=1}=0.
\end{equation}
Due to the Lemma 1 the conditions (\ref{vbc}) are consistent with
the time evolution
\begin{eqnarray}\label{tv}
&&D_t\pmatrix{\phi\cr\psi}=-4D^3_x\pmatrix{\phi\cr\psi}+6u
D_x\pmatrix{\phi\cr\psi}+3u_x \pmatrix{\phi\cr\psi}+\nonumber\\&&
+w \pmatrix{\alpha&0\cr0&-\alpha} \pmatrix{\phi\cr\psi}
\end{eqnarray}

According to the theorem above the Marchenko kernel which is not
scalar now but a vector-function satisfies the boundary value
problem
\begin{equation}\label{fvy}
\alpha D_y\pmatrix{F\cr\tilde F}=D_x^2\pmatrix{F\cr -\tilde F}
-D_z^2 \pmatrix{F\cr -\tilde F}
\end{equation}
with the conditions
\begin{equation}\label{fbc}
F-\tilde F|_{y=0}=0,\qquad F-\tilde F|_{y=1}=0.
\end{equation} Concentrate on the case of the degenerate
kernel \bb\label{5F} F(x,z,y,t)=\sum^N_{n=1} c_n
\exp((q_n^2-p_n^2)y -4(q_n^3+p_n^3)t+q_nz+p_nx). \ee By means of
the constraint $\tilde F(x,z,y,t)=F(z,x,y,t)$ one gets
\bb\label{5tF} \tilde F(x,z,y,t)=\sum^N_{n=1} c_n
\exp((q_n^2-p_n^2)y -4(q_n^3+p_n^3)t+q_nx+p_nz). \ee Substituting
the boundary conditions in the points $y=0$ and $y=1$ one gets
\begin{equation}\label{5inv}
c_n=c_{N-n},\quad q_n=p_{N-n},\quad p_{N-n}=\bar p_n, \quad
p_n^2-\bar p_n^2=2ik\pi.
\end{equation}
The last constraint shows that the quantity $p_n$ is described by
an entire and a real parameters $k\in{\bf Z}$ and $\beta_n\in
{\bf R}$ such that $$p_n=p_n(k)=\frac{k\pi}{2\beta_n}+i\beta_n$$
for $n=1,...N/2.$

One can replace each of the conditions (\ref{bc010}) (or both
simultaneously) by the condition $u_x+w|_{y=y_0}=0$ taking either
$y_0=0$ or $y_0=1$. Then the boundary conditions for the Lax pair
(\ref{vbc}) will be replaced by $\phi-\psi_x|_{y=0}=0$, or
respectively, $\phi-\psi_x|_{y=1}=0$. The
Gel'fand-Levitan-Marchenko kernel will satisfy instead of
(\ref{fbc}) the following boundary condition $F_x+\tilde
F_z|_{y=0}=0$, or respectively, $F_x+\tilde F_z|_{y=1}=0$.

{\bf Remark}. The problem is open how to apply the inverse
scattering transform method to the problems on the stripe.

\section{Cutting off constraint for discrete chains}

The well known Hirota equation is a reduction of the following
basic equation (see, for instance, \cite{zab})
\begin{equation}\label{hir}
(D_k-1)(\frac{t^{-k}_{u}t^{k}_{v}-t_{u}t_{v}}{tt_{uv}})=0,
\end{equation} Here $t=t(u,v,k)$ is the dependent variable.
Super- and sub-indices denote shifts with respect to discrete
arguments: $t^{-k}=t(u,v,k-1)$, $t^{k}=t(u,v,k+1)$,
$t_u=t(u+1,v,k)$, $t_v=t(u,v+1,k)$. In order to get Hirota
equation one should put the expression inside of the curly
brackets in (\ref{hir}) equal to $-1$. The Lax pair for the
equation (\ref{hir}) is given as pair of chains of Laplace
transformations
\begin{eqnarray}\label{laxu}
&&\psi_u=\psi^k+\frac{tt^{k}_{u}}{t_ut^k}\psi,\\ \label{laxv}
&&\psi_v=\psi+\frac{t^k_vt^{-k}}{t_vt}\psi^{-k},
\end{eqnarray}
for the following discrete hyperbolic equation
\begin{equation}\label{hyp1}
\psi_{uv}= \psi_{u}+\frac{t_vt_{uv}^k}{t^k_vt_{uv}}\psi_{v}-
\frac{tt_{uv}^k}{t^k_vt_{u}}\psi,
\end{equation}
where $\psi_u=\psi(u+1,v,k)$ and $\psi_v=\psi(u,v+1,k)$ etc.
(about discrete Laplace transforms see survey by I.A.Dynnikov,
S.P.Novikov, \cite{dyn}). The iterations of Laplace transform
above are enumerated by upper index. The operators forming these
transforms are $D_u$ and $D_v$. There is one more Lax pair for
the equation (\ref{hir}), which is based on the other kind of the
Laplace chain generated by operators $D_u^{-1}$ and $D_v^{-1}$
\begin{eqnarray}\label{laxu2}
&&\phi_{-v}=\phi^k+\frac{tt^{k}_{-v}}{t_{-v}t^k}\phi,\\
\label{laxv2}
&&\phi_{-u}=\phi+\frac{t^k_{-u}t^{-k}}{t_{-u}t}\phi^{-k},
\end{eqnarray}
serving the folloving hyperbolic equation
\begin{equation}\label{hyp2}
\phi_{uv}=\frac{t_ut_{v}^k}{t^kt_{uv}}\phi_{u}+
\frac{t_vt_{u}}{tt_{uv}}\phi_{v}-
\frac{t_ut_{v}^k}{t^kt_{uv}}\phi.
\end{equation}

Note that discrete hyperbolic equations admit two more pairs of
mutually inverse Laplace transforms based on the operators
$D_u^{-1}$, $D_v$ and $D_u$, $D_v^{-1}$. So totally they have four
kinds of Laplace transforms. Two pairs define as one could see
above two different Lax pairs. We show below that two others
define the cutting off constraint.

Proposition 2. Let the equations (\ref{hyp1}) and (\ref{hyp2}) be
connected by substitutions $$\psi= (aD_u+b)\phi,$$ then the
following boundary condition (cutting off constraint) is
satisfied $$(D_u-1) \frac{t_{uv}t-t_ut_v}{(t_v^k)^2}=0.$$ And the
substitution and its inversion are of the form
\begin{equation}\label{sub1}
\psi=\frac{t_{u}}{t^1}(1-D_u)\phi,\qquad
\phi=\frac{t_{-v}}{t^1}(D_v^{-1}-1)\psi.
\end{equation}

These substitutions are nothing else but the mutually inverse
Laplace transforms with operators $D_u$ and $D_v^{-1}$,
respectively. By replacing $u\leftrightarrow v$ one would get the
other kind of BC's. Let us reduce the chain (\ref{hir}) into a
segment $k\in(0, N)$ by setting the BC's at the ends $k=0$, $k=N$:
$$t(u,v,-1)=0;  \quad (D_u-1)
\frac{t_{uv}t-t_ut_v}{(t_v^k)^2}|_{k=N}=0;$$ Eigenfunctions
$\phi(u,v,k)$ and $\psi(u,v,k)$ are also restricted to the same
segment $k\in(0, N)$. At the left end the linear equations of the
Lax pair are cut evidently under BC. At the right end one has to
prescribe any of two equivalent equations (\ref{sub1}) taken at
the point $k=N$.

Proposition 3. Let the equations (\ref{hyp1}) and (\ref{hyp2}) be
connected by second order substitutions
\begin{eqnarray}\label{sub2}
&\psi&=(D_u-1)z(D_u-1)\phi, \nonumber\\
&\phi&=(D_v^{-1}-1)z(D_v^{-1}-1)\psi,
\end{eqnarray}
defined as compositions of two Laplace transforms, where
$\displaystyle{z={t_u\over t^1}}$, then the following boundary
condition is satisfied $$(D_u-1){t\over t_u^{-1}}=0.$$

The substitutions found allows one to cut up the infinite Lax
pairs and adopt them to restricted chains.

\section{Cutting off constraint for Hirota's equation}

Let us pass to the Hirota equation
\begin{equation}
T_uT_v-TT_{uv}=T^k_vT^{-k}_u\label{hiro2}.
\end{equation}
For this reduction the boundary conditions above can be
integrated under which they take (up to the point symmetries) the
following forms

i) $T(u,v+1,1)=T(u+1,v,-1);$

ii) $T(u,v,0)=T(u+1,v,-1).$

In the continual limit the Hirota equation approaches the Toda
chain (see, for instance, \cite{zab}), the boundary conditions
found above turn into cutting off constraint for the chain
corresponding to the Lie algebras of series $C_n$ and $B_n$. The
case corresponding to the series $A_n$ is well known, it has been
found years ago. But the case of $D_n$ is not found yet, it is
connected with rather labour consuming computations.

The discrete Toda chain (\ref{dtoda}) can also be studied by the
method above.

\end{document}